\documentclass{article}

\title{On an Alternative Cosmology}
\date{May 21, 1998}

\begin{document}
\maketitle

\subsection*{Introduction}
 The Standard Cosmological Model faces numerous problems
 (see, for example, [1-3]).
 The present work is an attempt to formulate an alternative
 cosmological concept to resolve the problems on a basis of
 a set of physical hypotheses.

\section{Hypotheses}

 The Grand Universe (GU) is an infinite multitude of Typical
 Universes (TUs).  TUs have limited volumes and are chaotically
 and uniformly, on average, dispersed in a Grand Universe Space
 (GUS).  Matter, in all forms, outside the TUs is a Grand Universe
 Background (GUB).  Our Home Universe (HU) is an example of a TU
 consisting of stars, black holes (if any), and other matter
 called a Typical Universe Background (TUB).  A Local Universe
 (LU) is the visible part of the HU.  Observed Cosmic Background
 Radiation (CBR) is a typical component of TUB.  A Local Grand
 Universe (LGU) is an imaginary part of the GU with a random
 sample of TUs.

\subsection*{Basic hypotheses in the GU concept}

\begin{enumerate}
\item The GU exists in the GUS, which is an infinite, flat 3-space.
 The GU is in a steady state.  It is a uniform and isotropic, on
 average, universe.  It has both baryon and electric symmetry.
 It has no absolute reference system (absolute "aether").
\item The intrinsic properties of GU matter are relativistic
 motion and self-recreation.
\item The GU exists in the form of matter and antimatter TUs,
 which evolve due to their interaction with the GUB.
\end{enumerate}

 The given hypotheses are in principle sufficient for a numerical
 simulation of GU matter behavior but seem to be redundant.
 In the author's opinion, a generalized matter transport theory
 must be developed for this purpose. Let us consider a statistical
 ensemble of GU matter under conditions formulated in hypothesis (1).
 Material objects are supposed to interact in direct collision
 and by gravitational forces while in flight.  Conventional physical
 processes must be taken into account.  Then all GU properties
 stated in hypotheses (2) and (3) are expected to be deduced from
 the theory.  In particular, a GU matter ensemble has to be
 characterized by a smooth, self-sustained mass distribution
 (equally for matter and antimatter) in a broad mass range,
 beginning with elementary free particles and ending with a TU
 itself.  The distribution function must be one of the main goals
 of the theory.  The problem is that any evolving TU, being a
 sub-ensemble, may appear to be more complex than the ensemble
 itself.  Further, we take the above set of hypotheses for granted
 to conjecture qualitative physical consequences and to draw a
 cosmological picture in the framework of the GU concept, as we
 call it.

\section{ Relativistic Motion of GU Matter}

 The postulated absence of absolute aether in GUS means that a
 coordinate-momentum distribution of GU matter must be a Lorentz
 invariant.  In other words, in any reference system it takes a form:
\begin{equation}
 f(x_1,x_2,x_3;p_1,p_2,p_3)=const 
\end{equation}

 This characterizes the chaotic motion of a relativistic photon
 or corpuscular gas [4].  In our case it applies to any material
 objects or gravitationally linked object systems as well.  The
 integral over momentum distribution in (1) diverges because we
 deal with an ideal model of non-interacting gas.   For interacting
 particles the distribution will have a smooth cut-off of upper
 energies in correspondence with limited energy density of GUB.

\section{TU Origin and Evolution. The Expanding Universe.}

 Relativistic motion of GU matter is a cause of a steady state and
 cyclic GU recreation in the processes of annihilation and pair
 creation.  One can imagine a probability chain of random object
 collisions resulting in a smooth mass distribution of separate
 bodies or gravitationally linked body systems.  Small but certain
 chances are given to an initial matter (antimatter) coagulant  to
 survive and grow to a "mature" TU age.  We may call the relevant
 process a TU embryo evolution.  As the TU evolves, it captures a
 large amount of cold matter from the GUB.  For this reason a TU
 space must store thermal radiation emitted by the cold matter.
 Yet, cold matter provides for the origination of dust clouds and
 stars.  Evolving stars converge some gravitational and nuclear
 energy into heat with a subsequent impact on TU properties.
 There is another reason for changing TU properties during its
 evolution.  A TU of greater mass is able to capture bodies of
 higher speed.  Eventually a TU becomes a relativistic system with
 growing instability.  It cannot collapse but may easily decay at
 any stage if its kinetic energy  happens to exceed its potential.
 A rough criterion of decay may be written as:
\begin{equation}
\frac{GM}{Rc^2(\gamma-1)}<1     
\end{equation}
 where: $M$-mass, $R$-mean radius, $\gamma $-effective Lorentz
 factor,$G$-gravitational con\\stant, $c$-speed of light.

 Thus, a TU is evolving due to its active interaction with GUB but
 inherent factors are essential.  An energy distribution of TU matter
 is different from that of GUB.  First, the mean energy is smaller. 
 Second, there must be a space-energy correlation: high speed objects
 move with higher radii.
 
 Keeping in mind this picture of TU evolution, let us consider our
 HU. Tracing back to red shifts,$z{\sim}10$, we can imagine the HU
 being about three orders more dense than at present, having a mass
 of roughly $M{\sim}10^{52}$kg and a radius $R{\sim}4*10^{25}$m,
 which is apparently close to criterion (2).  The most effective
 mechanism triggering the HU to decay might be an abrupt mass drop
 due to annihilation in a process of accidental "soft collision" with 
 a smaller antimatter TU.As a result, we observe the present Expanding
 Universe with the Hubble's expansion law.  Some features of the
 Expanding Universe are discussed below. 

\section{ Quasars}

 We have to explain several quasar features.  An effective volume
 of a quasar "engine" is about $10^{37} m^3$ with a power density
 of about $10^3 W  m^{-3}$.  Quasars came into existence in the last
 epoch of maximal z during what is thought to be the start of expansion.  
 Before that, a long period of HU pre-expansion evolution is supposed
 to have taken place.   Quasars may vary their power by about $50\%$
 in a short period of time.  Finally, they are apparently located in
 the center of galaxies.  One can start by recalling that annihilation
 is a process of potentially the highest power density possible in
 conventional physics.  We can assess $3*10^9$ nucleon pair
 annihilation in a unit volume per second for  a power rate of
 $1 W× m^{-3}$.  Suppose an "engine" volume is filled in with
 rock-like particles of a size $\alpha$ and density $\rho $. 
 Let particle clouds be mixed with corpuscular antimatter radiation
 with a flux $\varphi $.  Then an annihilation rate,$A$, is:
\begin{equation}    
 A ={\rho}{\alpha}^2{\varphi}   
\end{equation}

 Assuming
 $\rho=10^{-6}m^{-3},\alpha=10^{-1}m,{\varphi}=3*10^{20} m^{-2}s^{-1}$ 
 we have the required quasar power.  This rough estimate shows a
 physical feasibility of the annihilation nature of quasars. 
 Presumably they are active gravitating centers where exploding
 processes cause varying gas pressure and subsequently varying power.
   
 Gamma rays bursts having cosmological origin have been discovered
 some years ago.  We think the phenomenon has the same physical nature
 as quasars.  Antimatter objects constantly penetrating from the GUB
 cause random annihila\\tion pulses when colliding with HU stars. If
 so, a statistical analysis of gamma rays bursts should reveal a HU
 boundary.
  
\section{ Redshift Distribution}

 We assume a physical space to be Minkovski's space.  Its space-time
 scales are proportional to those of a conformally flat space taken
 in the original Freedman's solution for an expanding universe.  For
 simplicity, we neglect a peculiar velocity field and take a model of
 a uniformly expanding coherent gas of particles.  A gas system in a
 form of an expanding sphere has an initial effective radius $R(0)$. 
 In this case the special relativity theory gives us a velocity
 distribution of moving particles [5]:  
\begin{equation}     
 f(\beta)d\beta = Const*\gamma^4\beta^2d\beta   
\end{equation}
 for: $0\le\beta\le\beta_{max}$ where: $\beta_(max)$ is determined by
 $R(0)$.

 As usual,$\gamma=(1-\beta^2)^{-1/2}$ . 
        
 A velocity distribution (4) corresponds to a momentum distribution (1)
 if one takes into account that relativistic mass depends on velocity. 
 Hence, an initial chaotic distribution is consistent with a uniform
 HU expansion.  Neglecting a peculiar velocity field means that
 initial velocities are supposed to have radial alignment. 
     
 If we take a velocity distribution (4) as an approximate picture
 seen by a imaginary observer at the HU center and at the initial
 moment, $t=0$, we can easily find a corresponding redshift
 distribution $\varphi_i(z)$: 
\begin{equation}
 {\varphi}_i(z)=f({\beta}(z))\frac{d\beta}{dz}   
\end{equation}
 with a special relativity relation:    
\begin{equation}
 \beta(z)=\frac{(1+z)^2-1)}{(1+z)^2+1)}  
\end{equation}

 Then:  
\begin{equation}
 {\varphi}_i(z)=const*\frac{z^2(2+z)^2}{4(1+z)^3}       
\end{equation}

 In a process of expansion a deceleration influences velocity and
 corresponding redshift distributions making them dependent on time.
 A radial coordinate of some luminous, moving object is: 
\begin{equation} 
 r(t)=c\int_{0}^{t}\beta(t)dt                    
\end{equation}

 Suppose a light packet was emitted at emission moment te and
 registered at the moment of observation $t_0$.  Then an object
 velocity at $t=t_e$ is found from the equation:
\begin{equation}
 \int_{0}^{t_e}\beta(t)dt=(t_0-t_e)   
\end{equation}

 Therefore, a set of $t_e$ values and corresponding $\beta_e$ values is
 available for finding an emission velocity distribution
 $\varphi_0(\beta_e)$ and a corresponding redshift distribution
 $\varphi_0(z)$ observed at $t = t_0$.

\begin{equation}
 \varphi_0(\beta_e)=\varphi_i(\beta_i(\beta_e))\frac{d\beta_i}{d\beta_e}  
\end{equation}
 and 
\begin{equation}
 \varphi_0(z)=\varphi_0(\beta_e(z))\frac{d\beta}{dz}      
\end{equation}
 for: $0{\le}z{\le}z_{max}({\beta_e}_{max}))$ 

 The last expression concludes the solution of the redshift problem. 
 At small $z$:
\begin{equation}
 {\varphi}_0(z)=z^2 
\end{equation}
 and, the smaller  the ${\beta_e}_{max}$, the broader the range.
 During  the  present  epoch  redshift   counts show  the  dependence
 (12) at  least  up  to $z\approx 0.25$ [1].  This is evidence that
 appreciable time passed since the HU started expanding.  Detailed
 numerical analysis in the GU concept framework must reveal a good
 deal of knowledge about our HU.

 The Standard Model failed to predict redshift counts.  The reason
 for this failure, as well as criticism of the Standard Model, is
 not a subject of this work.

\section{``Black Matter'' and CBR}

 As discussed in previous sections, a TU captures cold matter from
 the GUB or creates its own in a form of star remnants during its
 evolution.  This is baryon cold matter, referred to in the Standard
 Model as "Black Matter".  It is uniformly distributed, on average,
 in TU space.  Cosmic background radiation (CBR) is a thermal
 radiation partially in relaxation with cold matter.  The surface
 temperature of cold matter in the HU is 2.7 K at present.  In
 general this temperature is determined by the local energy balance. 
 Apparently, a balance of absorbed and emitted energy is negative,
 and cold matter is cooling in a process of expansion.  Its
 temperature seems to be inversely proportional to an expansion
 factor.  A detailed study of the relationship between CBR and cold
 matter in the GU concept framework must clarify the situation.  

\section{Cosmic Rays}

 The GU concept gives a natural explanation of high energy cosmic
 rays observed from Earth: an ultrarelativistic GUB gas must be
 their source.  Therefore, cosmic rays are a result of GUB gas
 transformation and moderation in a process of transport through
 an HU medium.  Qualitatively we may assume a logarithmic energy
 loss being proportional to a time of GUB particle travel in an
 HU medium:
\begin{equation} 
 \frac{d\varepsilon}{\varepsilon}=-{\alpha}dt   
\end{equation}
 and a Poisson probability to reach an observer:        
\begin{equation}
 w(t)dt=\frac{dt}{T}exp(-\frac{t}{T})   
\end{equation}
 where $\alpha$, $T$ are transport model parameters.

 Then an observed spectrum is:
\begin{equation}
 F_0({\varepsilon})=Const*{\varepsilon}^{-(1+1/\alpha T)}
\end{equation}

 which is formed in a broad energy range
 ${\varepsilon}<{\varepsilon}_i$ $({\varepsilon}_i$ is the initial
 energy of relativistic GUB particles).  For ${\alpha}T=2/3$ (which
 is a logarithmic energy loss per effective particle life time) we
 have a spectrum close to that observed.

 A GUB gas is baryon symmetric but most primary antiparticles must
 have been annihilated in the transport process, excepting those in
 the upper energy range.  Unfortunately, measurement and
 identification of cosmic rays at superhigh energies is an extremely
 difficult task.

 Both CBR and cosmic rays are part of a TUB, therefore they play a
 role of absolute aether participating in an expansion process.
 However, boundary effects must cause some departures from Hubble's
 law.  This means that an observer who is not in the center of the HU
 must see an asymmetry in angular distribution.  Some anisotropic
 effects are really found, but higher measurement precision is needed
 for their interpretation. 

\section{ Some Other Cosmological Issues}

 There are Standard Model problems of singularity, flatness, horizon, 
 and "univer-\\sal aether" which have a quite general and, to some
 extent, philosophical character.  Fortunately, these do not arise in
 the alternative cosmology concept.  Other specific physical problems
 are related to the common question of whether Hubble's time is long
 enough comparing with the time required for star evolution and
 recycling, galaxy formation and clustering, large-scale structuring,
 nucleosynthesis, and so on.  The alternative concept does not impose
 time constraints on this sort of problems because  all processes are
 considered in a time scale that includes the GU recreation cycle and
 the pre-expansion stage of the HU.

 The alternative concept predicts physical boundary of any TU, the HU
 in particular.  Therefore, the HU must have a rotation motion in a
 GUS, and there must be some evidence of it.  Subsequent anisotropic
 effects in the HU picture will depend on our location. 

\subsection*{Conclusion}

 The concept qualitatively explains available observational data for
 our Home Universe but predictions concerning the whole Universe seem
 to have the same level of likelihood.  A general transport theory is
 needed to investigate in detail the suggested cosmological framework. 

\subsection*{Acknowledgements}

 I would like to thank Prof. N. Rabotnov for fruitful discussion of
 the problem.  I also wish to acknowledge Prof. J. Peebles for
 encouraging contacts.
 
\subsection*{References}

{\bf 1.} P.J.E. Peebles. {\it Principles of Physical Cosmology.}
 Princeton University Press. 1993.

{\bf 2.} {\it Critical Dialogues in Cosmology.}  In celebration of
 the 250th anniversary of Princeton University.  24-27 June, (1996).
 World Scientific Press.

{\bf 3.} M. Rees.  {\it Before the Beginning.}  Helix Books,
 Addison-Wesley Press, 1997.

{\bf 4.} J.L. Synge.  {\it The Relativistic Gas.}  North-Holland
 Publishing Company, Amsterdam.  1997.

{\bf 5.} V.A. Fok. {\it Theory of Space, Time, and Gravity.}
 (in Russian) Published in Moscow.  Teoriya prostranstrva, vremeni
 i tyagoteniya.  Gosizdatelstvo teckniko-teoreticheskoi literaturi.
 Moskva, 1955.

\end{document}